\documentclass[fleqn,10pt]{wlscirep}
\usepackage[utf8]{inputenc}
\usepackage[T1]{fontenc}
\usepackage{multirow}
\usepackage{multicol}
\usepackage{amsmath}
\usepackage{amsfonts}
\usepackage{amssymb}
\usepackage{graphicx}
\usepackage{graphicx,amssymb,amsmath,yhmath}
\usepackage{url}
\usepackage{caption}
\usepackage{commath}
\usepackage{comment}
\usepackage{aligned-overset}
\usepackage{stackengine}
\usepackage{tabu}
\usepackage{float}
\usepackage{commath}
\usepackage{authblk}
\usepackage{todonotes}
\usepackage{verbatim}
\usepackage{xcolor}
\usepackage{siunitx}
\usepackage{titlesec}
\title{Multi-class motion-based semantic segmentation for ureteroscopy and laser lithotripsy}

\author[1,*]{Soumya Gupta}
\author[1,2]{Sharib Ali}
\author[3]{Louise Goldsmith}
\author[3]{Ben Turney}
\author[1,*]{Jens Rittscher}
\affil[1]{ Institute of Biomedical Engineering and Big Data Institute, Department of Engineering Science, University of Oxford, Oxford, UK}
\affil[2]{Oxford NIHR Biomedical Research Centre, Oxford, UK}
\affil[3]{Department of Urology, The Churchill, Oxford University Hospitals NHS Trust, Oxford, UK}
\affil[*]{soumya.gupta@eng.ox.ac.uk and jens.rittscher@eng.ox.ac.uk}

\begin{abstract}
Kidney stones represent a considerable burden for public health-care systems, with the total health-care expenditure for kidney stones exceeding US \$ 2 billion annually in the USA alone. Ureteroscopy with laser lithotripsy has evolved as the most commonly used technique for the treatment of kidney stones. Automated segmentation of kidney stones and laser fiber is an important initial step to performing any automated quantitative analysis of the stones, particularly stone-size estimation, that can be used by the surgeon to decide if the stone requires further fragmentation. Factors such as turbid fluid inside the cavity, specularities, motion blur due to kidney movements and camera motion, bleeding, and stone debris impact the quality of vision within the kidney and lead to extended operative times. To the best of our knowledge, this is the first attempt made towards multi-class segmentation in ureteroscopy and laser lithotripsy data. We propose an end-to-end convolution neural network (CNN) based learning framework for the segmentation of stones and laser fiber. The proposed approach utilizes two sub-networks:~I) HybResUNet, a hybrid version of residual U-Net, that uses residual connections in the encoder path of U-Net to improve semantic predictions, and II) a DVFNet that generates deformation vector field (DVF) predictions by leveraging motion differences between the adjacent video frames which is then used to prune the prediction maps. We also present ablation studies that combine different dilated convolutions, recurrent and residual connections, atrous spatial pyramid pooling and attention gate model. Further, we propose a compound loss function that significantly boosts the segmentation performance in our data. We have also provided an ablation study to determine the optimal data augmentation strategy for our dataset. Our qualitative and quantitative results illustrate that our proposed method outperforms state-of-the-art methods such as UNet and DeepLabv3+ showing an improvement of 5.2\% and 15.93\%, respectively, for the combined mean of DSC and JI in our {\textit{\textit{in vivo}}} test dataset. We also show that our proposed model generalizes better on a new clinical dataset showing a mean improvement of 25.4\%, 20\%, and 11\% over UNet, HybResUNet, and DeepLabv3+, respectively, for the same metric. 
\end{abstract}
\begin{document}
\flushbottom
\maketitle
\thispagestyle{empty}

\section{Introduction}
\label{sec:introduction}
Kidney stones are a common urological disease that affects about 12\% of the world population. It has a recurrence rate of 10\% after one year, 50\% over a period of 5-10 years and 75\% over 20 years in most patients~\cite{alelign2018kidney}. Kidney stones, also known as renal calculi, are formed when crystal forming substances separate from the urine and accumulate inside the upper urinary tract, kidney, ureter or bladder~\cite{alelign2018kidney}. Typically, stones larger than 5~\si{mm} can result in a blockage in the ureter, inducing severe pain in the abdomen and the lower back~\cite{miller2007management}. Ureteroscopy has evolved into a minimally invasive routine technique for treatment of a number of urological conditions such as urolithiasis, strictures and hematuria~\cite{reddy2010ureteroscopy}.

Technological advancements have led to the development of low-cost single-use endoscopes with improved flexibility and image quality. The data used in this study has been acquired using single-use LithoVue$^\text{TM}$ scope and Lithovue Elite scope (Boston Scientific). The procedure involves inserting a long flexible ureteroscope into the urinary tract through urethra passing through the bladder and then into the ureter and kidney to access the kidney stones. The scope also has a working channel through which tools like laser fiber can be inserted to perform laser lithotripsy, i.e., stone fragmentation using laser energy. Based on the size, location and composition of stone, the surgeon decides if the stone requires dusting or fragmentation and sets the laser settings accordingly ~\cite{holmium}. Fragmented stones are either left in place to clear out by themselves over time or extracted using a special wire basket. The surgeon tries to carefully target the stone centrally, rather than peripherally, in order to limit the excess heat generated in the confined spaces of the kidney or ureter~\cite{holmium}. A ureteral stent is usually inserted into the ureter by the surgeon for easy passage of the residual stone debris and fragments. When stones are larger than the diameter of the ureter, an additional surgery is required. In order to avoid such discomfort to patients and assist clinicians to perform targeted laser treatment, estimating the size and location of the kidney stones is important.

\begin{figure}[t!]
    \centering
    \includegraphics[width=0.7\textwidth]{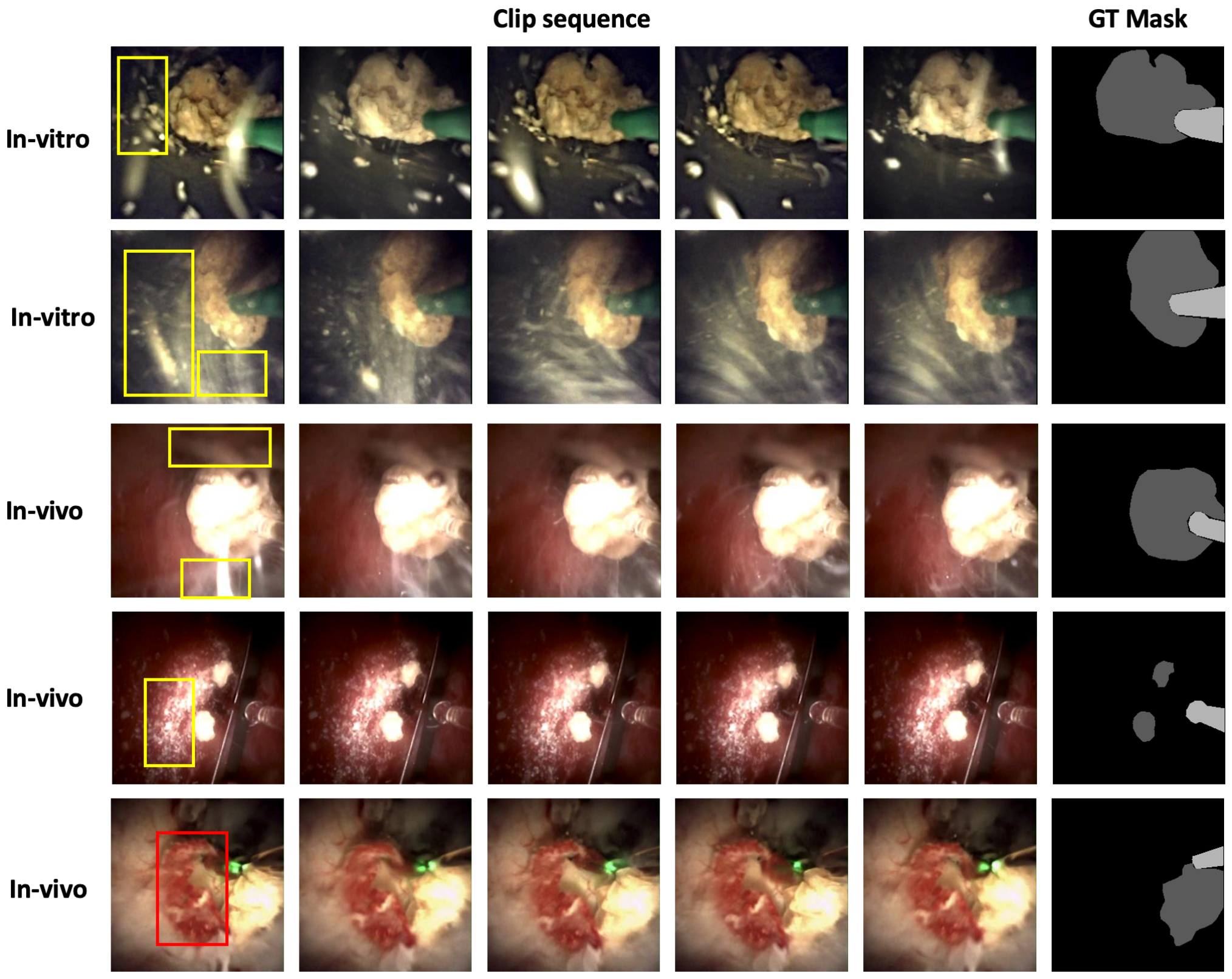}
    \caption{Exemplary images for the \textit{in vitro} and \textit{in vivo} clip sequences with corresponding ground truth masks showing stone and laser fiber. Stone debris and blood are highlighted in yellow and red rectangles (on left), respectively.}
    \label{dataset}
\end{figure}

Automated segmentation is the primary step to perform any analysis of the stone fragments and laser fiber. However, compared to standard radiology, very little work has been published to address the problem of automated segmentation in ureteroscopy video. The ureteroscopy and laser lithotripsy data is significantly different and challenging as compared to the other endoscopy datasets in numerous aspects. The ureteroscopy video has a small-field-of-view and the signal quality is affected by stone debris originating from stone fragmentation that obscures the vision in the kidney, making it difficult for surgeons to perform the stone-treatment procedure efficiently, thereby adding to the treatment time. Intra-operative bleeding can also occur during ureteroscopy due to the continuous application of laser energy, intra-renal pressure and the trauma caused to the walls of the ureter~\cite{de2019complications}. In addition to the aforementioned challenges, the segmentation task becomes even more complex due to the motion blur arising from unavoidable kidney movements and camera motion, specular highlights, dynamic background, varying illumination conditions, artefacts from the turbid fluid inside the target cavity~\cite{rosa2011algorithm}, and high variation in the size, shape and composition of the stone. Sample images from our \textit{in vitro} and \textit{in vivo} datasets are shown in Figure~\ref{dataset} wherein the stone debris and blood are highlighted in yellow and red rectangles, respectively. 

To the best of our knowledge, multi-class segmentation that delineates both stones and laser fiber in ureteroscopy video has not been addressed before. Our previous work \cite{soumyaSPIE, soumyaISBI} focuses on the segmentation of stone fragments in ureteroscopy video. However, segmentation of laser fiber is also important considering the fact that inaccurate laser targeting can result in undesirable excess heat and extended operative times. Here, we have experimented with various combinations of residual and recurrent connections, dilated convolutions, ASPP layers and attention module to identify the semantic segmentation network with competitive performance. To further improve the network performance, we make use of the temporal information in sub-sequences by incorporating  a sub-network called DVFNet that leverages motion between the adjacent frames to compute end-to-end deformation vector field (DVF) predictions. This motion information is then used to prune the segmentation map obtained from the semantic segmentation network, resulting in a context-aware edge enhanced multi-class segmentation. The main contributions made in this work can be summarized as follows:
\begin{itemize}
    \item A novel end-to-end CNN-based learning framework with residual connections that leverages motion between image pairs to overcome inevitable challenges of motion blur, stone debris and other artefacts, and provide real-time multi-class segmentation of both stone and laser fiber in ureteroscopy and laser lithotripsy dataset. 
    
    \item A novel compound loss function is proposed that outperforms traditional loss functions on ureteroscopy and laser lithotripsy dataset. 
    
    \item Experimental validations on diverse and challenging {\textit{in vitro}} and {\textit{in vivo}} ureteroscopy datasets demonstrate the effectiveness of our proposed multi-class segmentation approach.

\end{itemize}
We compare our proposed method with different state-of-the-art (SOTA) methods to exhibit competitiveness of the approach. We conduct a generalizability test of our network on an unseen second test dataset to measure the robustness compared to the SOTA methods. Finally, we provide an extensive ablation study that validates our network choices and data augmentation strategies (see \textbf{Supplementary material}). The type of dataset used in this study has not been well-explored in literature, and is significantly different and challenging as compared to the other endoscopy datasets (exemplary images are provided in Fig.~\ref{dataset}).

\section{Related work}
This section builds on recent advances in semantic segmentation and image registration. Of particular relevance are those segmentation and registration methods that have been developed for endoscopy imaging.

\subsection{Semantic segmentation}
Here, we first outline different methods that have been proposed for segmentation of kidney stones. We then discuss deep learning methods used for segmentation of various organs in endoscopy imaging with a particular focus on encoder-decoder networks. 

\textbf{Segmentation of kidney stones.} Several methods such as
Region indicator with Contour segmentation (RICS)~\cite{tamilselvi2012segmentation}, 
modified watershed segmentation~\cite{tamilselvi2012modified},
and squared euclidean distance method~\cite{tamiselvi2013segmentation} have been implemented for the detection and segmentation of kidney stones in ultrasound (US) images. Some studies have also explored techniques such as intensity, location and size based thresholds~\cite {thein2018image}, Fuzzy C-means clustering followed by level set~\cite{akkasaligar2017kidney},
and CNN~\cite{langkvist2018computer}
for detection and segmentation of kidney stones in CT images. Rosa et al.~\cite{rosa2011algorithm} proposed a region growing algorithm for renal calculi segmentation on ureteroscopy images.  However, such approaches require user’s intervention to define seed pixel, similarity criterion and a stopping criterion which is very challenging to determine due to the nature of variability in kidney stones. Gupta et al. proposed an optical flow based segmentation technique for binary segmentation of stone fragments in ureteroscopy~\cite{soumyaSPIE}. Gupta et al. also presented an end-to-end convolutional network that leveraged motion differences between adjacent frames to further improve the segmentation of stones in ureteroscopy and laser lithotripsy data~\cite{soumyaISBI}. 

\textbf{Segmentation of various abnormalities in endoscopy.} Handcrafted features have been applied for the segmentation and detection of various abnormalities such as bleeding~\cite{tuba2017algorithm}, polyps~\cite{prasath2017polyp},  ulcers~\cite{yuan2015saliency} and  tumor regions~\cite{alizadeh2014segmentation} in endoscopy videos.
Various deep-learning based strategies on automated segmentation of bleeding zones in wireless capsule endoscopy (WCE) have been proposed~\cite{jia2017bleeding,ghosh2018bleeding}. 
Ali et.al \cite{ali2020translational} presented a comprehensive analysis of various approaches that were submitted to 
EAD2020 challenge for artefact detection and segmentation and EDD2020 challenge for disease detection and segmentation. A multi-scale context guided deep network based on FCN was proposed by Wang et.al~\cite{wang2020multi} for lesion segmentation in endoscopy images of Gastrointestinal (GI) tract.  Jha et. al~\cite{jha2020real} presented a deep learning based approach for real-time detection, localisation and segmentation of polyps in colonoscopy.

\textbf{Encoder-decoder models based on recurrent and residual connections.} The efficacy of encoder-decoder based CNNs for semantic segmentation has been well established for a large number of applications. U-Net~\cite{ronneberger2015u} is one of the most popular encoder-decoder models that combines high level semantic information with low level details via skip connections to boost the segmentation accuracy in biomedical datasets. Zhang et al.~\cite{zhang2018road} improvised U-Net performance by adapting a deep residual U-Net architecture (DeepResUNet) that combined the strengths of deep residual learning~\cite{he2016deep} and U-Net architecture~\cite{ronneberger2015u}. Peretz et. al~\cite{peretz_amar_2019} reported that replacement of all feed-forward blocks with residual units tend to make the network complicated and results in overfitting. They then suggested a hybrid version of U-Net called HybResUNet for brain tumor segmentation wherein residual blocks are only used in the encoding path of the U-Net. A U-Net with residual connections has also been recently proposed for instrument segmentation in endoscopic images~\cite{isensee2020or}. Alom et al. proposed a Recurrent residual U-Net (R2-UNet)~\cite{alom2018recurrent} that uses recurrent convolutional layers with residual connectivity for improved medical image segmentation. 

\textbf{Encoder-decoder models based on dilated convolutions.} Several studies have proposed that replacing conventional convolutions in CNN models with dilated convolutions helps to aggregate multi-scale contextual information without losing resolution and significantly improves the network performance~\cite{yu2015dilation,hamaguchi2018dilation,piao2019dilation}. Hamaguchi et. al~\cite{hamaguchi2018dilation} claimed that increasing dilation factors in any network tends to increase sparsity of kernel and fails to aggregate local features. They then proposed a novel architecture for segmentation of small object instances in remote satellite imagery by first increasing the dilation factors and then decreasing them. Atrous spatial pyramid pooling (ASPP) that involves parallel atrous convolutions with different dilation rates has also boosted the accuracy of detection and segmentation of objects at different scales~\cite{Deeplabv3+}. Recently, a variant of Residual UNet called ResUNet-a~\cite{resunetDiakogiannis} was proposed where atrous convolutions and pyramid scene parsing pooling was incorporated in the network to improve the segmentation accuracy. 

\textbf{Encoder-decoder models based on attention gates.} Attention mechanisms when incorporated in neural networks have proven to be effective in highlighting only the relevant activation during training and are computationally efficient. 
A novel attention gate model~\cite{oktay2018attention} was proposed and integrated into the standard U-Net to highlight relevant features that are passed through the skip connections and actively suppress the irrelevant and noisy responses in skip connections for pancreas segmentation. This Attention U-Net~\cite{oktay2018attention} was also integrated with R2-UNet~\cite{alom2018recurrent} for improved segmentation~\cite{leejunhyun2019}. Jha et.al~\cite{jha2019resunet++} proposed ResUNet++ that took advantage of residual units, ASPP and attention units to provide improved segmentation of colorectal polyps. 

Although the field of endoscopy imaging has been well explored, computer vision work in the field of ureteroscopy and laser lithotripsy is still at its infancy. The ureteroscopy and laser lithotripsy data is significantly different and challenging as compared to the other endoscopy datasets in numerous aspects that include: significant amount of stone, blood and other debris that obscure the target; dynamic background; high variation in the appearance, size, shape and composition of stone;specular highlights; high motion blur; and additional image artefacts from turbid fluid inside the target cavity. In this study, we have tried to overcome some of these challenges by experimenting with various combinations of residual and recurrent connections~\cite{peretz_amar_2019,alom2018recurrent,resunetDiakogiannis}, ASPP~\cite{Deeplabv3+}, dilated convolutions~\cite{hamaguchi2018dilation} and attention gate model~\cite{oktay2018attention} that have been integrated to the base network U-Net to obtain an improved multi-class semantic segmentation in ureteroscopy and laser lithotripsy data.

\subsection{Image registration}
Several studies have shown the image registration and segmentation are complementary tasks, meaning the features learned in image registration can be used to improve the segmentation result as well~\cite{qin2018joint, mahapatra2018joint}.  To tackle limitations of simple rigid transformations, deformable image registration (DIR) methods are used for most works in medical image analysis~\cite{cao2018deformable, ali2019conv2warp}. Some non-learning based approaches for DIR such as diffeomorphic Demons~\cite {vercauteren2009diffeomorphic}, 
HAMMER~\cite{shen2002hammer} and FNIRT~\cite{andersson2008fnirt} have gained tremendous popularity. However, Traditional methods of image registration are iterative, time consuming. and often fail when there is a lot of variation in appearance between the source and target images.

\textbf{Supervised learning based DIR methods.} Several supervised DIR approaches have been developed in the literature. 
A CNN based regression model was developed for brain MR images to directly learn mapping between the source and target images to their corresponding DVFs~\cite{cao2018deformable}. Yang et al.~\cite{yang2017quicksilver} also proposed a network to predict deformable registration followed by its refinement using a correction network for brain MR images. Such neural network methods, however, rely on strong supervision for training. Use of supervised methods are majorly limited by the fact that they require ground-truth deformation vector fields (DVFs) for model training which is difficult to obtain, especially in case of medical datasets. 

\textbf{Unsupervised learning based DIR methods.} Unsupervised DIR methods have gained tremendous success and popularity over the recent years. 
In order to handle large non-linear deformations, Vos et al.~\cite{de2019deep} used B-spline for transformation and interpolation for the predicted deformation fields and presented results on much complex datasets. However, B-spline does not pass through all data points and can often lead to large interpolation errors~\cite{ali2019conv2warp}. Sharib et al.~\cite{ali2019conv2warp} in 2019 presented an unsupervised end-to-end CNN framework for image registration that used a bicubic Catmull-Rom spline resampler to reduce the errors in the resampling of deformation fields. They also added a series of deformable convolutional filters to better capture complex deformations. 

Building on the work of Ali et al.~\cite{ali2019conv2warp}, our proposed framework involves a sub-network called DVFNet that leverages motion between the adjacent frames to compute end-to-end deformation vector field (DVF) predictions which are then used to prune the results of our semantic segmentation network.

\section{Materials and Method}
This section starts with the description of the dataset used in our study and is followed by the details on our proposed framework for multi-class segmentation in ureteroscopy and laser lithotripsy. 

\subsection{Materials and experimental set-up }

The dataset (\textit{in vitro} and \textit{in vivo}) used in this study has been acquired using the Boston Scientific LithoVue scope. The \textit{in vitro} dataset was acquired under controlled settings wherein laser lithotripsy of four different human kidney stones was individually performed inside a container with irrigation fluid flowing through it.
The invivo dataset used in this study was provided by the Oxford University Hospitals and Boston Scientific.Sub-sequences containing intense lasering and stone movement were extracted and clip sequences containing 5 adjacent frames of these sub-sequences were used. One frame from each clip sequence (1 out-of 5) was manually labelled using the following  three class labels: stone fragments, laser fiber, and background. The VGG Image Annotator (VIA) tool~\cite{vgg} was used to obtain a ground truth mask for each clip sequence as shown in Figure~\ref{dataset}. The annotations used in this study were performed by a PhD student and independently verified by two experts, one of which was a senior urologist. 

\begin{figure}[t!]
    \centering
    \includegraphics[width=0.9\textwidth]{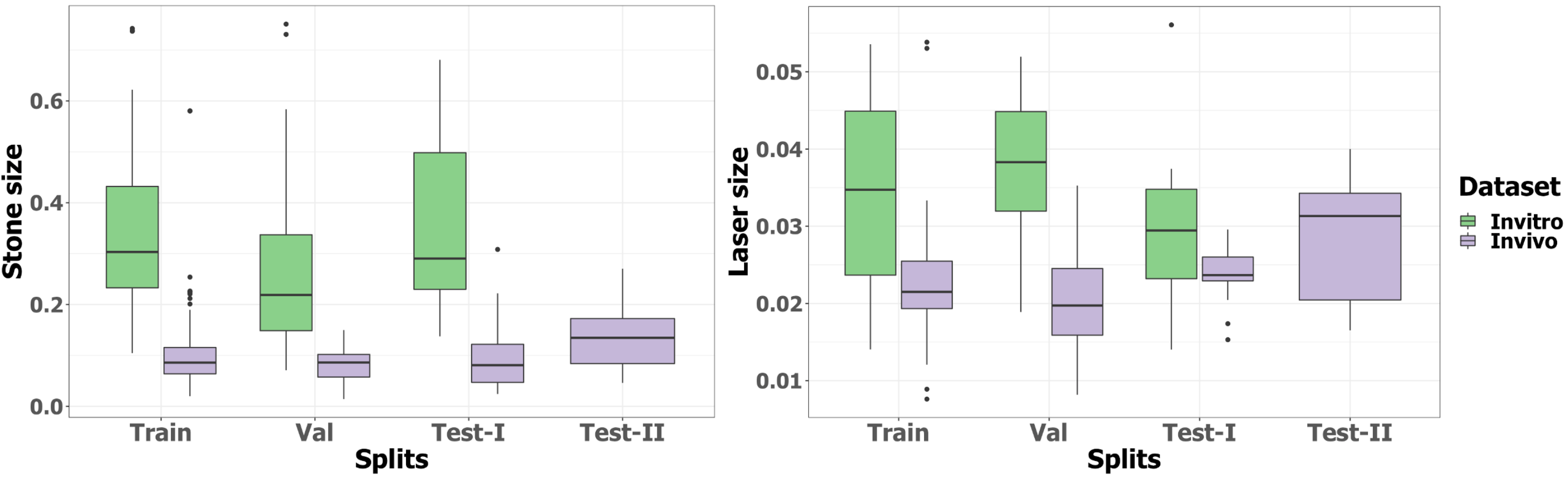}
    \caption{Box-plots showing size distribution of stone and laser class across training, validation and test sets in the \textit{in vitro} and \textit{in vivo} datasets. It can be seen that size distribution of kidney stones is different across train, validation and test sets, indicating that they come from different videos. The large variability between the \textit{in vitro} and \textit{in vivo} datasets is also clearly evident, explaining why we ran experiments independently for the \textit{in vitro} and \textit{in vivo} datasets.}
    \label{size_dis}
\end{figure}

The dataset was randomly split into train (60\%), validation (20\%) and test (20\%) sets for both \textit{in vivo} and \textit{in vitro} data. An overview of the dataset configuration has been shown in Table~\ref{config}. For a better understanding of the datasets, box-plots showing relative size distribution of stone and laser class across training, validation and test sets in the \textit{in vitro} and \textit{in vivo} datasets, respectively, has been shown in Figure~\ref{size_dis}. As evident from Table~\ref{config} and Figure~\ref{size_dis}, high standard deviation of the stone indicates high variability of kidney stones in each dataset. It can also be seen that the mean of stone is different for train, validation and test sets, indicating that they come from different videos. Most importantly, it is important to note a large variability between the \textit{in vitro} and \textit{in vivo} datasets. We therefore conducted independent experiments for \textit{in vitro} and \textit{in vivo} setting. In case of \textit{in vivo}, we have included an extra dataset (Test-II) that consists of unseen new samples and is used in the final part of this study to test the generalizabiliity of our model as compared to existing SOTA approaches. All image samples in this study were each resized to $256\times 256$ pixels in RGB format to train the networks. Networks were trained with a batch size of 2 on NVIDIA Quadro RTX 6000 for 100 epochs using Adam optimizer with a learning rate of 1e$^{-3}$, initial decay rates were set to default 0.9 and 0.999 for estimation of the first and second moments of gradient respectively, with validation performed after every epoch. 

\begin{table}[]
\begin{center}
\begin{tabular}{ c|c|c|c|c|c } 
\hline
\multirow{2}{*}{\bf Dataset} & \multirow{2}{*}{\bf Split}&
\multicolumn{2}{c|}{\bf Number of samples}
&  \multicolumn{2}{c}{\bf Distribution ($\mu \pm \sigma$)}\\
\cline{3-6}
&  &  Single & Sequence ($\times$5)& Stone & Laser \\
\hline 
&Train&52&260 &$0.3570\pm0.1702$ & $0.0345\pm0.0107$\\ 
\cline{2-6}
\textit{In vitro} & Validation&18&90 &$0.3517\pm0.1635$  &  $0.0295\pm0.0099$\\ 
\cline{2-6}
& Test-I&18&-- &$0.2858\pm0.1975$  & $0.0369\pm0.0101$ \\ 
\hline 
&Train&92&460 &$0.1084\pm0.0865$ & $0.0222\pm0.0067$\\ 
\cline{2-6}
\textit{In vivo} &Validation&32&160 & $0.0836\pm0.0337$ &  $0.0205\pm0.0069$ \\ 
\cline{2-6}
&Test-I &30&-- &$0.0949\pm0.0628$ & $0.0240\pm0.0033$ \\ 
\cline{2-6}
&Test-II &20&-- &$0.1384\pm0.0566$ & $0.0277\pm0.0076$ \\ 
\hline
\end{tabular}
\end{center}
\caption{Overview of the train, validation and test splits for the vitro and \textit{in vivo} experiments. Mean ($\mu$) and standard deviation ($\sigma$) for each stone and laser annotations in each split is also provided.
\label{config}}
\end{table}

\subsection{Method}

Our proposed framework utilizes two sub-networks: {HybResUNet}, which is a hybrid version of residual U-Net that uses residual connections in the encoder path only;and {DVFNet} that leverages motion between adjacent frames to compute end-to-end deformation vector fields. This motion information obtained from DVFNet is used to prune the segmentation mask obtained from the semantic segmentation network resulting in an improved multi-class segmentation. The entire framework is designed as an end-to-end CNN model that optimizes our proposed compound loss function.  In this section, the semantic segmentation network module, DVFNet and the compound loss function are presented.

\subsubsection{HybResUNet}
\textbf{Base network.} The first sub-network of our segmentation framework is an encoder-decoder based network called HybResUNet~\cite{peretz_amar_2019}.It is basically a 9-level deep U-Net architecture wherein residual blocks are used instead of traditional feed-forward units in the contracting (encoder) path as shown in Figure:~\ref{network}. This is because replacement of all feed forward units in conventional encoder-decoder network with residual blocks increase the network complexity and tend to overfit the training data~\cite{peretz_amar_2019}. Each of the four residual units in the encoder consists of repeated application of 
two $3\times3$ convolutions, followed by a Batch Normalization (BN) and a Rectified Linear Unit (ReLU). These convolutions are followed by an addition of output to its initial input as in residual units and a $2\times\ref{aug}$ max pooling operation with a stride of 2 for downsampling. 
The decoder in the HybResUNet uses transposed and regular convolutions to gradually increase the image size while reducing the number of features. The network also consists of skip connections that circumvent the information loss during downsampling by concatenating the output obtained from each residual block with the output of transposed convolution from the up-scaled features at the decoder layers.
Each of these concatenations is further followed by sequential application of two regular convolutions. For a fair comparison, other networks namely U-Net~\cite{ronneberger2015u}, DeepResUNet~\cite{zhang2018road}, and R2-UNet~\cite{alom2018recurrent} are all implemented as 9 level deep architecture to obtain the best performing base network for our dataset. 

\textbf{Dilations, ASPP and Attention gate.} To further improve our base network, we have tried incorporating dilated convolutions, Atrous Spatial Pyramid Pooling (ASPP) and an attention gate mechanisms~\cite{oktay2018attention}. Replacing conventional convolutions in CNN models with dilated convolutions is known to improve the aggregation of  multi-scale contextual information without losing resolution~\cite{yu2015dilation,hamaguchi2018dilation, piao2019dilation}. Hamaguchi et. al~\cite{hamaguchi2018dilation} introduced a novel architecture for segmentation of small object instances in remote satellite imagery by first increasing the dilation factors and then decreasing them. Inspired by this idea, we have empirically obtained a series of dilation rates that work best for our data: $[1,2,3,4,3,2,1,2,1]$. These networks also consist of an Atrous Spatial Pyramid Pooling (ASPP) module at the end of the encoding path with an output stride (ratio of the input image size to the output feature map size) of 16 and six parallel convolutions with dilation rates $[1, 2, 4, 8, 16, 32]$. To sum up, the experiments labeled with ASPP in Table~\ref{quant_network} (and Supplementary material Table~\ref{quant_network-supplem}) comprise of a series of dilated convolutions and an ASPP module at the end of the encoding path. To leverage the attention gate mechanism~\cite{oktay2018attention}, we add attention gates to the skip connections just before the concatenation operation (refer to Att in Table~\ref{quant_network} and Supplementary material Table~\ref{quant_network-supplem})). Such mechanism suppresses the propagation of irrelevant and noisy responses in the network. We further tried to incorporate ASPP, series of dilated convolutions and attention gate all together to observe if this improves the segmentation accuracy (refer to Att-ASPP in Table~\ref{quant_network-supplem}). 

\subsubsection{DVFNet}
Building on the work of Ali et. al~\cite{ali2019conv2warp}, our DVFNet is also based on an encoder-decoder architecture where the parameters of the spline resampler are learnt from training data. It consists of total 12 layers, that include three linear convolutional layers, two average pooling layers and two deformable convolutional layers in the encoder. Each of these convolutional layers is combined with Batch Normalization (BN) and exponential linear unit (ELU) as shown in Figure~\ref{network}. The decoder layer consists of a Catmul-Rom spline resampler to rescale the obtained DVF from the encoder, which is then further resampled with two additional layers that consist of a convolution layer, an ELU activation and the spline resampler. The final DVF obtained at $0^{th}$ scale is then applied on the original image to obtain the corresponding warped image $I_{warp}$.
\subsubsection{Loss function}
Although commonly used, the cross entropy loss not differentiate between easy (correctly-classified) and hard (misclassified) samples causing the easily classified negatives in hard samples to compromise the majority of the loss and dominate the gradient~\cite{lin2017focal}. Focal loss (FL), which is an improved version of CE loss was introduced by Lin et al.~\cite{lin2017focal}.The Focal Loss is defined as:
\begin{equation}{\label{eq:FL}}
    L_{FL} = - (1-\hat{p}_{y})^{\gamma} \log (\hat{p}_{y})
\end{equation}
where, y $\in$ \{0, ..., C-1\} is an integer class label (C denotes the number of classes), $\hat{p}$ = \{($\hat{p}_{0}$,...,$\hat{p}_{C-1}$)\} $\in$ [0,1]$^{C}$ is a vector representing an estimated probability distribution over the C classes and $\gamma\geq0$ is the tunable focusing parameter (set to default value of 2) wherein higher the $\gamma$, the higher is the rate at which easy-to-classify examples are down-weighted. FL is a dynamically scaled CE loss where the scaling factor $\gamma$ decays to zero as confidence in the correct class increases (Eq.~(\ref{eq:FL})). Intuitively, the scaling factor  $\gamma$ automatically down-weights the easy examples and forces the model to focus on hard examples~\cite{lin2017focal}. 

Bokhovkin et. al~\cite{bokhovkin2019boundary} proposed a novel loss function that is essentially a differentiable surrogate of a metric accounting accuracy of boundary detection. Let's say $y_{pd}$ and $y_{gt}$ represent the binary map predicted by a neural network and ground truth map, respectively for arbitrary class c for an image. The boundaries $y_{gt}^{b}$ and $y_{pd}^{b}$ can then be defined as:
\begin{align}
    y_{gt}^{b} = pool(1-y_{gt},\theta_\circ) - (1-y_{gt}) \quad \text{and} \nonumber \\
    y_{pd}^{b} = pool(1-y_{pd},\theta_\circ) - (1-y_{pd}),
\end{align}
\noindent{where} (1-$y_{gt,pd}$) refers to the inversion of any pixel of the map and pool(·, ·) denotes a pixel-wise max-pooling operation to the inverted binary map with a sliding window of size, $\theta_\circ$ set to 3. The euclidean distances between pixels to boundaries requires computation of a supporting map which is the map of extended boundary given by $ y_{gt}^{b,ext} = pool( y_{gt}^{b}\theta)$ and $y_{pd}^{b,ext} = pool( y_{pd}^{b}\theta)$, where $\theta$ set to 5. The precision $P^{c}$ and recall $R^{c}$ can then be defined as: 

\begin{align}
    P^{c} = \frac {sum(y_{pd}^{b} \circ y_{gt}^{b,ext})}{sum(y_{pd}^{b})} \\
    R^{c} = \frac {sum(y_{gt}^{b} \circ y_{pd}^{b,ext})}{sum(y_{gt}^{b})},
\end{align}
\noindent{where} `$\circ$' denotes the pixel-wise multiplication of two binary maps and {sum}(.) refers to the pixel-wise summation of a binary map. The reconstructed boundary metric, $B^{c}$ is averaged over all classes and is then used to formulate the loss function, $L_{boundary}$ that can be defined as: 

\begin{equation}
    L_{boundary} = 1-{B^c} \text{   with boundary metric  } {B^c} = \frac{2P^{c}R^{c}}{P^{c}+R^{c}}
\end{equation}
where $P^{c}$ and $R^{c}$ refer to the precision and recall. Bokhovkin et. al~\cite{bokhovkin2019boundary} performed a comparative analysis of their proposed boundary loss with various loss functions such as IOU loss, Dice loss and Sensitivity-Specificity (SS) loss. In the first part of our study where we use non-sequence data, we aim to find the best performing baseline network by using a compound loss function that combines this boundary loss with SOTA Focal loss. For the second part of this study where we have integrated DVFNet for motion estimation, we propose to use our extended novel compound loss function which is a combination of Focal loss, Boundary loss, Cross-correlation loss between warped and target image, and smoothness constraint~\cite{icnet} on the predicted deformation fields.
The cross-correlation loss is computed between the warped images ($I_{{warp}_{1\leftarrow 3}}$ and $I_{{warp}_{3\leftarrow 5}}$) with their corresponding source image ($I_{1}$ and $I_{3}$), respectively and is given by:
%
\begin{equation}
    L_{sim} =  L_{sim1}(I_{1},I_{{warp}_{1\leftarrow 3}}) +  L_{sim2}(I_{3},I_{{warp}_{3\leftarrow 5}})
\end{equation}
i.e.,
\begin{equation}
    L_{sim}= \frac{1}{2N} \sum \bigg(\frac{I_{1}(x) - \mu _{1}}{\sqrt{{\sigma _{1}}^{2}+{\epsilon}^{2}}} - \frac{I_{{warp}_{1\leftarrow 3}}(x) - \mu _{warp_{1\leftarrow 3}}}{\sqrt{{\sigma _{warp_{1\leftarrow 3}}}^{2}+{\epsilon}^{2}}}\bigg)^{2} +
    \frac{1}{2N} \sum \bigg(\frac{I_{3}(x) - \mu _{2}}{\sqrt{{\sigma _{2}}^{2}+{\epsilon}^{2}}} - \frac{I_{{warp}_{3\leftarrow 5}}(x) - \mu _{warp_{3\leftarrow 5}}}{\sqrt{{\sigma _{warp_{3\leftarrow 5}}}^{2}+{\epsilon}^{2}}}\bigg)^{2}
\end{equation}
where, $\mu$ and $\sigma$ are the mean and standard deviation, N is the total number of pixels and $\epsilon$ = 10$^{-3}$ to avoid division by zero. The estimated deformation vector fields (DVFs) can be locally smoothed using a smoothness constraint on its spatial gradients. Further, the smoothness constraint on the estimated deformation vector fields ($DVF_{1 \leftarrow 3}$ and $DVF_{3 \leftarrow 5}$) can be expressed as: 
\begin{align}
    L_{smo}= \sum (\: \norm{\nabla \mbox{ $DVF_{1 \leftarrow 3}$}}_{2}^{2} + \: \norm{\nabla \mbox{ $DVF_{3 \leftarrow 5}$}}_{2}^{2} ),
\end{align}
where $\nabla$ indicates the gradient of flow fields and $\norm{.}$ indicates its $L_{2}$ norm. 

Finally, our proposed compound loss function can be formulated as:
\begin{equation}
    L = L_{FL}+\alpha L_{boundary}+ \beta L_{sim}+ \zeta L_{smo},
    \label{fullloss}
\end{equation}
where $\alpha$,  $\beta$, and $\zeta$ are the hyper-parameters used to balance the contribution from the boundary loss, similarity loss and smoothness loss, respectively with initial values set to 1, 0.5 and 1, respectively. 

\subsubsection{Proposed framework}
\begin{figure}[ht!]
    \centering
    \includegraphics[width=1.0\textwidth]{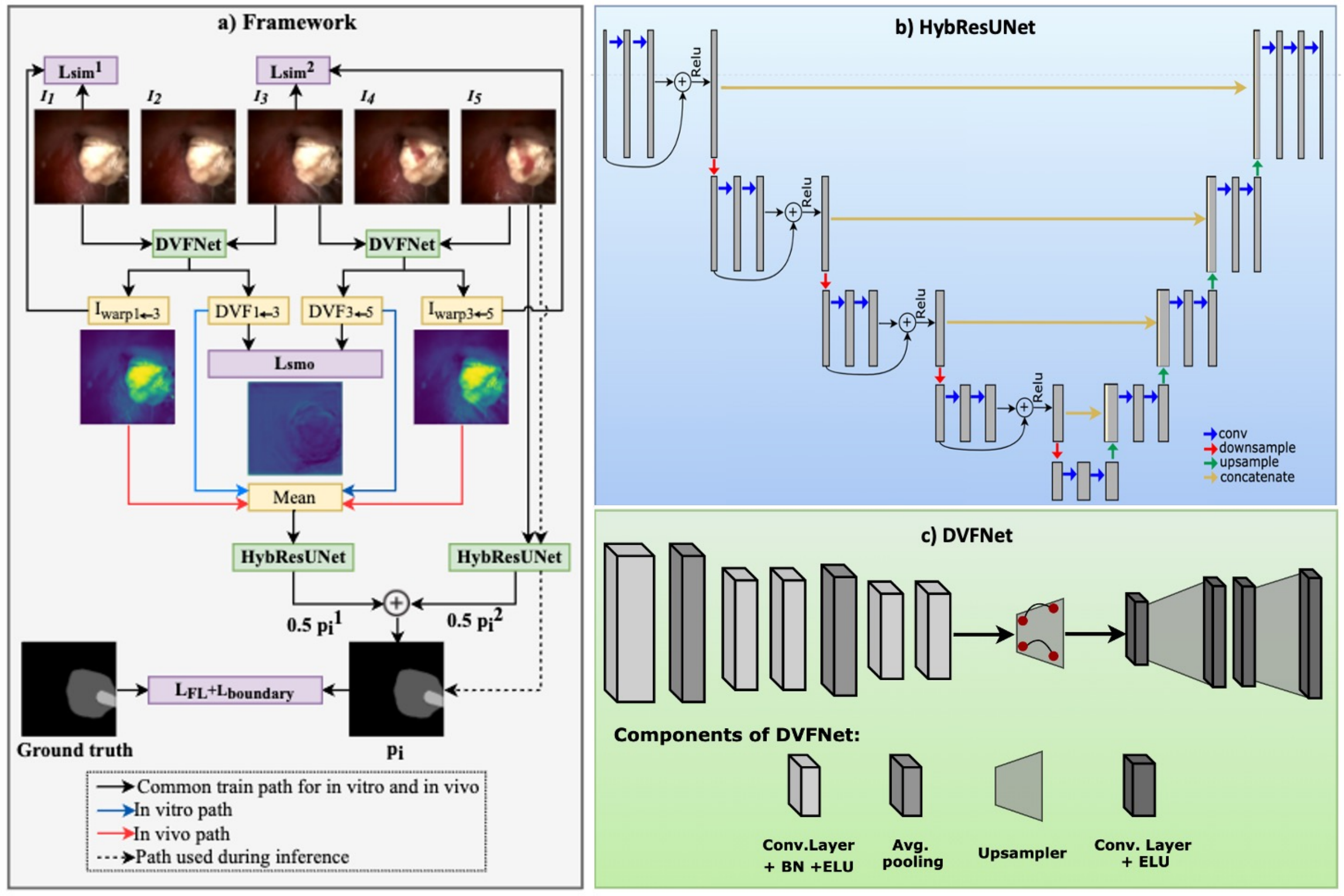}
    \caption{\textbf{Proposed framework:} a) Overall proposed semantic segmentation framework illustrating both training and inference paths presented in this study. Here, DVFNet utilizing image pairs are only used for training shown by solid path while the learned weights of HybResUNet are used during frame wise inference indicated by the dotted line. b) Network representation of the HybResUNet used in this work. It is a 9 level deep U-Net architecture with residual connections in the encoder path and c) an encoder-decoder DVFNet architecture used to compute deformation vector fields (DVFs) between image pairs that are used to prune the segmentation results obtained from the HybResUNet.}
    \label{network}
\end{figure}

Our proposed framework is shown in Figure~\ref{network}. Each clip sequence comprises of five images, $I_{1}$ to $I_{5}$ wherein image pairs ($I_{1}$, $I_{3}$) and ($I_{3}$, $I_{5}$), respectively, are provided as an input in grayscale format to different DVFNet networks during training. Each DVFNet then computes a Deformation Vector Field (DVF) map ($DVF_{1\leftarrow 3}$ and $DVF_{3\leftarrow  5}$) and their corresponding warped image ($I_{{warp}_{1\leftarrow 3}}$ and $I_{{warp}_{3\leftarrow  5}}$). The obtained DVFs are locally smoothed via smoothness constraint $L_{smo}$ on its spatial gradients. Further, Normalized Cross Correlation (NCC) is used as a similarity metric to minimize shape differences between the obtained warped images ($I_{{warp}_{1\leftarrow 3}}$ and $I_{{warp}_{3\leftarrow 5}}$) and their corresponding source images $I_{1}$ and $I_{3}$, respectively, as shown in the Figure~\ref{network} ($Lsim^1$ and $Lsim^2$ correspond to these losses). 

In case of the \textit{in vitro} dataset, the mean of the two DVFs ($DVF_{1\leftarrow 3}$and $DVF_{3 \leftarrow 5}$) is fed as an input to the semantic segmentation network HybResUNet to obtain the first semantic map, $p_{i}^{1}$. In contrast for the \textit{in vivo} data, the mean of the warped images ($I_{{warp}_{1\leftarrow 3}}$ and $I_{{warp}_{3\leftarrow  5}}$) is fed to the HybResUNet network to obtain the first semantic map, $p_{i}^{1}$. In Figure~\ref{network} colored arrows (blue, red) represent the corresponding paths for both \textit{in vitro} and \textit{in vivo} data for these choices. 
A second semantic map, $p_{i}^{2}$ is obtained by feeding the fifth input image $I_{5}$ in the RGB format to another HybResUNet. Finally, the two semantic maps are averaged to obtain a final map, $p_{i}$. The network then optimizes the final output semantic map by minimizing a combined loss function represented in Eq.~(\ref{fullloss}). DVFNet part of the framework is only used during network training (indicated by solid path in Figure~\ref{network}) while the learned weights of the HybResUNet are used during frame-wise inference (indicated by dotted path in Figure~\ref{network}).

\section{Results}
Before comparing our proposed compound loss function with the SOTA focal loss~\cite{lin2017focal} we outline our data-augmentation strategy.
Subsequently quantitative and qualitative results of the proposed networks against existing SOTA networks on our laser lithotripsy and ureteroscopy datasets are presented. Finally, we provide a generalizability study of our proposed framework followed by a quantitative comparison of inference time of the networks. 

\subsection{Data augmentation}

Data augmentation techniques such as flips, random crops, rotate and color jittering have been commonly used for deep learning in medical imaging. In this study, we first intend to determine the optimal augmentation choices by studying the effect of different strategies on segmentation accuracy. Initially, we performed 8 training experiments for both {\textit{in vitro}} and {\textit{in vivo}} datasets, where each experiment involved training HybResUNet on one of the aforementioned augmentation types.  Refer to the \textbf{Supplementary material Table~\ref{aug-settings}} for the list of augmentation techniques with their corresponding settings used in our study. Dice similarity coefficient (DSC) was recorded for each experiment for the stone and laser class (refer to the \textbf{Supplementary material Table~\ref{aug}}). It can be seen that the \textit{Random Brightness Contrast(RBC)} and \textit{Equalize} transformation improve the segmentation accuracy in the {\textit{in vitro}} datasets as compared to no augmentation scenario(refer to the  \textbf{Supplementary material Table~\ref{aug}}). On the other hand, \textit{RBC} and \textit{Contrast Limited Adaptive Histogram Equalization (CLAHE)} both provide a higher DSC compared to no augmentation in the {\textit{in vivo}} datasets (refer to the \textbf{Supplementary material Table~\ref{aug}}). The difference in the results of the {\textit{in vitro}} and {\textit{in vivo}} can be majorly attributed to the difference in background appearance between the two datasets (refer to Figure~\ref{dataset}). Further experiments in this study have therefore used \textit{RBC+Equalize} and \textit{RBC+CLAHE} for augmentation of the {\textit{in vitro}} and {\textit{in vivo}} datasets, respectively.  

\subsection{Loss function}
As detailed in Section 3.2.3, we use a novel compound loss function that is a combination of focal loss, boundary loss, cross-correlation loss and a smoothness loss for improved segmentation performance in our study. Table~\ref{lossfn} presents a quantitative comparison of the segmentation results obtained when a U-Net was trained with the focal Loss vs when trained with a combination of other losses in case of both \textit{in vitro} and \textit{in vivo} test datasets. It can be observed that the combination of focal loss and boundary loss improved the mean DSC by nearly 5.2 \% and 9.5 \% in case of \textit{in vitro} and \textit{in vivo} datasets, respectively. It can also be seen that incorporating similarity loss and smoothness loss to this combination of focal and boundary loss further boosted the mean DSC by 2.6 \% in case of \textit{in vitro} and 2.3 \% in case of the \textit{in vivo} dataset. It is also worth noting that when combined losses are used instead of focal loss, the DSC for the laser class is boosted by a higher margin as compared to the boost in the stone class. 
\begin{table}[h!]
\vspace{0.2in}
\small{
\begin{center}
\begin{tabular}{ c|c|c|c|c|c|c } 
\hline
\multirow{3}{*}{\bf Loss function} & \multicolumn{6}{c}{\bf DSC}\\
\cline{2-7}
& \multicolumn{3}{c|}{\bf \textit{In vitro}} & \multicolumn{3}{c}{\bf \textit{In vivo}}\\
\cline{2-7}
& Stone & Laser  & \bf{Mean} & Stone & Laser & \bf{Mean}\\
\hline
\hline
$L_{FL}$ & 0.8544 & 0.7643 & 0.8094& 0.7257&	0.6991& 0.7124 \\
$L_{FL}$+$L_{boundary}$ & 0.8631&0.8401 &0.8516& 0.7948	&0.7657&0.7803\\
$L_{FL}$+ $L_{boundary}$+$L_{sim}$+$L_{smo}$  & 0.8892 & 0.8582 &0.8737& 0.7825 &	0.8144 & 0.7985\\
\hline
\end{tabular}
\end{center}
}
\caption{Comparison of loss functions (applied on U-Net) showing accuracy improvement with compound loss as opposed to state-of-the-art focal loss\label{lossfn}}
\end{table}

\subsection{Quantitative results}
We have evaluated our proposed method and compared it with other existing SOTA approaches using standard computer vision metrics such as Dice similarity coefficient (DSC), Jaccard index (JI), positive predictive value (PPV) and senstivity in Table~\ref{quant_network}. The details on the test datasets are provided in Table~\ref{config}. In Table~\ref{quant_network} below, we have established a quantitative comparison of our proposed framework against SOTA network architectures. Our \textbf{Supplementary  material Table~\ref{quant_network-supplem}} presents an ablation study for integration of dilations, ASPP and attention gate in our network. 

\begin{enumerate}[label=(\Roman*)]
    \item  \textbf{Base network}: SOTA methods are compared as base network of choice for our proposed assembled network. For this case, it can be observed that the HybResUNet provided a significantly higher DSC and JI with mean of  0.8434 for \textit{in vitro} and 0.7822 for the \textit{in vivo}, higher overall PPV and sensitivity were also seen as compared to other baseline networks included in the experiment (U-Net, DeepResUNet, R2UNet and DeepLabv3+).
    
    \item \textbf{Base network with DVFNet (with DVF)}: In this set of experiments, we propose to incorporate DVFNet together with the best base network in (I), i.e., HybResUNet. For this network, the mean of $DVF_{1\leftarrow 3}$ and  $DVF_{3\leftarrow 5}$ is fed as input to the first HybResUnet as shown in Figure:~\ref{network}. In case of \textit{in vitro}, it can be seen that it significantly improved the mean of DSC and JI of both HybResUNet and Att-HybResUNet by nearly 1.1\% and 4.2\%, respectively. It can also be seen that the DVFNet further increased the overall PPV and sensitivity of most networks in this set. Although DVFNet (with DVF) can be seen to improve the stone segmentation of HybResUNet in the \textit{in vivo}, it showed no overall improvement in the segmentation results.   
    
    \item \textbf{Base network with DVFNet (with warped images)}: This sixth set of experiments involved incorporation of DVFNet (with warped image) together with the best base network {HybResUNet} and its derivatives. Here, warped image corresponds to the case when mean of $I_{{warp}_{1\leftarrow  3}}$ and $I_{{warp}_{3\leftarrow  5}}$ is fed to the input of the first HybResUnet as shown in Figure:~\ref{network}. In case of the \textit{in vitro}, DVFNet (with warped image) can be seen to improve the mean of DSC and JI of Att-HybResUNet and Att-ASPP-HybResUNet by 1\% and 1.63\%, respectively. However, the overall performance is still lower than that obtained from the use of DVFNet (with DVF). For the \textit{in vivo} data, it can be observed that DVFNet (with warped image) improved the mean of DSC and JI of all networks: HybResUNet, ASPP-HybResUNet, Att-HybResUNet, Att-ASPP-HybResUNet by nearly 1.63\%, 2.1\%,0.78\% and 2.23\%, respectively. In addition to this, DVFNet (with warped image) can be seen to increase the sensitivity of all networks in this set.
\end{enumerate}
\begin{table}[]
\vspace{0.2in}
\small{
\begin{center}
\resizebox{\textwidth}{!}{%
\begin{tabular}{ |c|cc|c|c|c|c|c|c|c|c|c| }
\hline
\multirow{2}{*}{\bf Dataset} &
\multicolumn{2}{c|}{\multirow{2}{*}{Method}}&
\multicolumn{2}{c|}{\bf DSC}
&  \multicolumn{2}{c|}{\bf JI} &
 \multirow{1}{*}{\bf Mean} &
\multicolumn{2}{c|}{\bf PPV}
&  \multicolumn{2}{c|}{\bf Sensitivity} \\
\cline{4-7}

\cline{9-12}
& & & Stone & Laser  & Stone & Laser& \bf{(DSC, JI)} & Stone & Laser & Stone & Laser\\
\hline 
\multirow{13}{*}{\bf \textit{In vitro}}&\multirow{5}{*}{\bf I}& UNet~\cite{ronneberger2015u} & 0.8521	&	0.8388	&	0.7736	&	0.7787	&	0.8108	&0.9155	&	0.8393	&	0.8479	&	0.8408\\
&& HybResUNet~\cite{peretz_amar_2019} & 0.8838	&	0.8724	&	0.8143	&	0.8031	&	{0.8434}	&	0.9123	&	0.8792	&	0.8914	&	0.8780\\
 && DeepResUNet~\cite{zhang2018road} & 0.8828	&	0.8276	&	0.8054	&	0.7541	&	0.8175	&	0.9239	&	0.8609	&	0.8732	&	0.8065\\
&& {R2-UNet}~\cite{alom2018recurrent} & 0.8215	&	0.8061	&	0.7344	&	0.7368	&	0.7747	&	0.8855	&	0.8380	&	0.8280	&	0.8013\\
&& DeepLabv3+(Resnet-50)~\cite{Deeplabv3+}	&0.8004	&	0.8446	&	0.6982	&	0.7717	&	0.7787	&0.8411	&	0.8626	&	0.8110	&	0.8452\\
\cline{2-12}
&\multirow{4}{*}{\bf II}& {HybResUNet+DVFNet (with DVF)}\textsuperscript{\textdagger}&\bf{0.9068}&	0.8702&	\bf{0.8413}&	0.7923&	0.8527&	0.9299&	0.851&	0.904&	0.9069	\\
&&{ASPP-HybResUNet+DVFNet (with DVF)}\textsuperscript{\textdagger}&0.8253	&	0.8278	&	0.725	&	0.7572	&	0.7838	&	0.8244	&	0.832	&	0.8847	&	0.832	\\
(Test-I)&&\bf{Att-HybResUNet+DVFNet (with DVF)}\textsuperscript{\textdagger}&0.8872	&	\bf{0.9004}	&	0.8180	&	\bf{0.8263}	&	\bf{0.8580}	&	0.9136	&	0.8891	&	0.8919	&	0.9227	\\
&&{Att-ASPP-HybResUNet+DVFNet (with DVF)}\textsuperscript{\textdagger}&0.8238	&	0.8475	&	0.738	&	0.7809	&	0.7976	&	0.862	&	0.8326	&	0.8585	&	0.874	\\
\cline{2-12}
&\multirow{4}{*}{\bf III}& {HybResUNet+DVFNet (with warped image)}\textsuperscript{\textdagger}&0.8801	&0.8708	&0.7985	&0.7898	&0.8348	&0.8899&	\bf{0.8955}&	0.8931	&0.8697	\\
&&{ASPP-HybResUNet+DVFNet (with warped image)}\textsuperscript{\textdagger}&0.8295	&	0.8758	&	0.7447	&	0.8066	&	0.8142	&	0.8165	&	0.8556	&	0.8768	&	0.9118	\\
&&{Att-HybResUNet+DVFNet (with warped image)}\textsuperscript{\textdagger}&0.8899	&	0.8393	&	0.8211	&	0.7711	&	0.8304	&	\bf{0.9534}	&	0.8312	&	0.8626	&	0.8536	\\
&&{Att-ASPP-HybResUNet+DVFNet (with warped image)}\textsuperscript{\textdagger}&0.886	&	0.8769	&	0.8089	&	0.7984	&	0.8426	&	0.8898	&	0.8548	&	0.9113	&	0.9233	\\

\hline
\hline
\multirow{13}{*}{\bf \textit{In vivo}}&\multirow{5}{*}{\bf I}&UNet~\cite{ronneberger2015u} & 0.8129	&	0.7974	&	0.7025	&	0.7043	&	0.7543	&	0.8312	&	0.8054	&	0.8374	&	0.8137\\

&& HybResUNet~\cite{peretz_amar_2019} & {0.8339}	&	0.8241	&	\bf{0.7381}	&	0.7328	&	{0.7822}	&	0.8462	&	0.864	&	0.8525	&	0.8214\\

 && DeepResUNet~\cite{zhang2018road} & 0.8214	&	0.7851	&	0.714	&	0.7023	&	0.7557	&	0.8167	&	0.8084	&	0.8543	&	0.7783\\

& &{R2-UNet}~\cite{alom2018recurrent} & 0.7734	&	0.7678	&	0.6536	&	0.6636	&	0.7146	&	0.758	&	0.7875	&	0.8328	&	0.7979\\

&& DeepLabv3+(Resnet-50)~\cite{Deeplabv3+}	&0.7653	&	0.7144	&	0.6438	&	0.62	&	0.6859	&	0.7573	&	0.7287	&	0.8375	&	0.7111\\

\cline{2-12}
&\multirow{4}{*}{\bf II}
&{HybResUNet+DVFNet (with DVF}\textsuperscript{\textdagger}&\bf{0.8347}	&0.7834	&0.7357	&0.6965	&0.7626&	0.8429&	0.7748	&0.8473	&0.8409	\\
&&{ASPP-HybResUNet+DVFNet (with DVF)}\textsuperscript{\textdagger}&0.8115	&	0.8048	&	0.7127	&	0.7034	&	0.7581	&	0.8148	&	0.7826	&	0.843	&	0.8385	\\
(Test-I)&&{Att-HybResUNet+DVFNet (with DVF)}\textsuperscript{\textdagger}&0.7996	&	0.8017	&	0.6914	&	0.7144	&	0.7518	&	0.7506	&	0.8352	&	\bf{0.9026}	&	0.8300	\\
&&{Att-ASPP-HybResUNet+DVFNet (with DVF)}\textsuperscript{\textdagger}&0.8072	&	0.8374	&	0.7036	&	0.7397	&	0.7720	&	0.7881	&	0.7979	&	0.8668	&	\bf{0.8975}	\\
\cline{2-12}
&\multirow{4}{*}{\bf III}&
\bf{HybResUNet+DVFNet (with warped image)}\textsuperscript{\textdagger}&0.8203	&	{0.8568}	&	0.7158	&	\bf{0.7878}	&	\bf{0.7952}	&	0.8226	&	\bf{0.8894}	&	0.8562	&	0.8487	\\
&&{ASPP-HybResUNet+DVFNet (with warped image)}\textsuperscript{\textdagger}&0.7992	&	\bf{0.8658}	&	0.6911	&	0.7852	&	0.7853	&	0.7817	&	0.8634	&	0.8682	&	0.877	\\
&&{Att-HybResUNet+DVFNet (with warped image)}\textsuperscript{\textdagger}&0.8183	&	0.8389	&	0.7156	&	0.7558	&	0.7822	&	0.8032	&	0.8798	&	0.8713	&	0.837	\\
&&{Att-ASPP-HybResUNet+DVFNet (with warped image)}\textsuperscript{\textdagger}&0.8016	&	0.8478	&	0.6918	&	0.7662	&	0.7769	&	0.8062	&	0.8276	&	0.8382	&	0.8784	\\
\hline
\end{tabular}
}
\end{center}
}
\caption{Quantitative comparison of proposed network architectures against existing approaches on our ureteroscopy and laser lithotripsy test sets (Test-I). The published networks have been accordingly cited and our experiment networks have been indicated by a \textdagger superscript. \label{quant_network}}
\end{table}

\subsection{Qualitative results}
\begin{figure}[t!]
    \centering
    \includegraphics[width=0.78\textwidth]{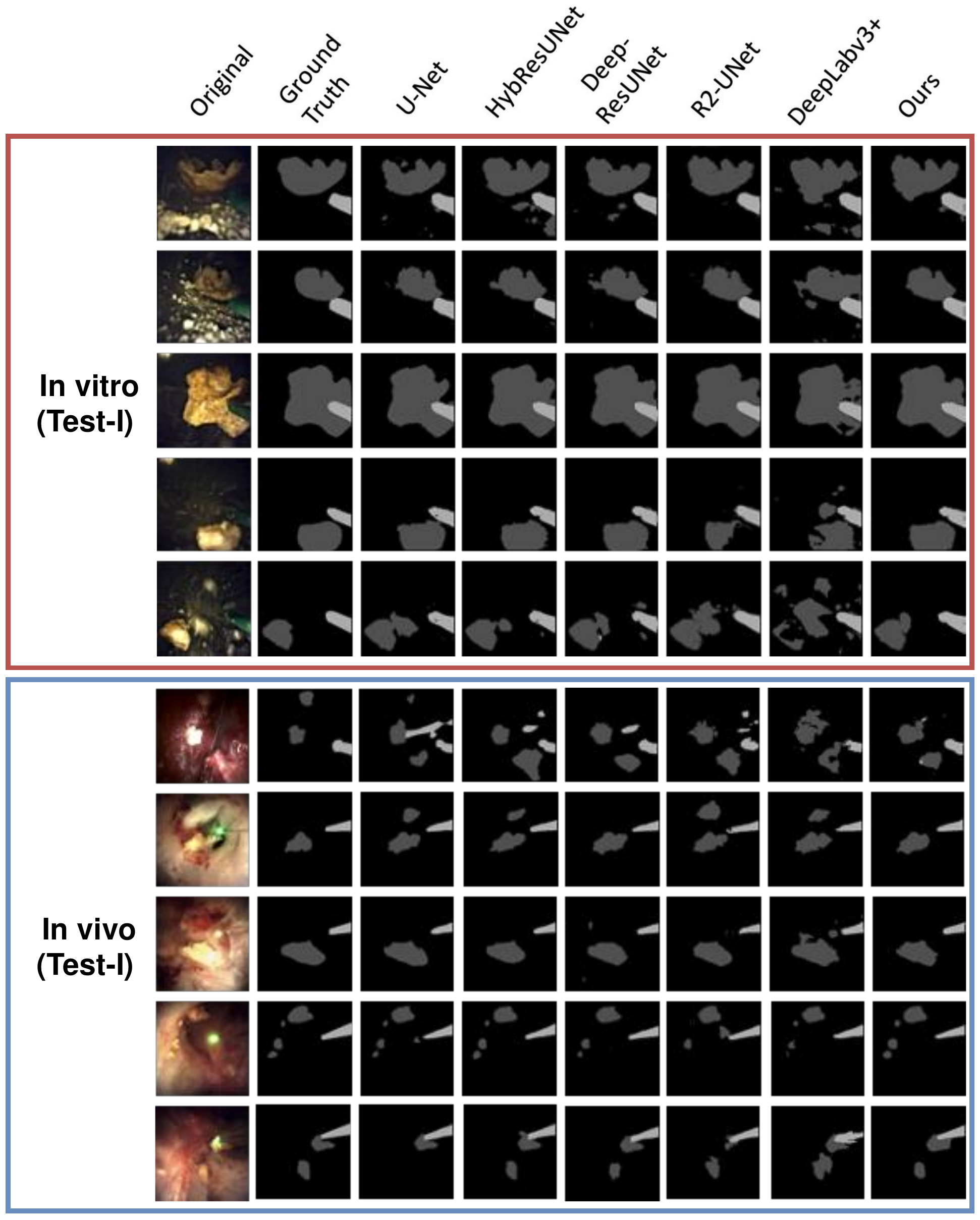}
    \caption{Qualitative analysis of our proposed  method (Att-HybResUNet+DVFNet (with DVF) for the \textit{in vitro} and HybResUNet+DVFNet (with warped image) for the \textit{in vivo}) against existing SOTA methods methods on our ureteroscopy and laser lithotripsy test sets (Test-I). Each row shows a test image, followed by its ground truth segmentation mask (showing laser fiber and stone), followed by SOTA approaches: UNet, HybResUNet, Deep-ResUnet, R2-UNet and DeepLabv3+, and finally our proposed model which is Att-HybResUNet+DVFNet (with DVF) for the \textit{in vitro} and HybResUNet+DVFNet (with warped image) for the \textit{in vivo}.}
    \label{qualitativeresults}
\end{figure}

In this section, we have presented the qualitative results of segmentation obtained from our proposed framework as opposed to the ground truth and other SOTA approaches in both \textit{in vitro} and \textit{in vivo} cases. Figure:~\ref{qualitativeresults} demonstrates that our model outperforms the existing approaches by overcoming the effect of debris and providing a more accurate delineation of stone and laser fiber in case of both \textit{in vitro} and \textit{in vivo} datasets. As evident in Figure:~\ref{qualitativeresults}, it can be clearly seen that laser fiber is nearly segmented well by all models except for some difficult frames like first image in the \textit{in vivo} dataset wherein it is only our model that is able to clearly segment the laser. It can also be observed from Figure~\ref{qualitativeresults} that the existing approaches are not able to provide a clear segmentation of stone in most \textit{in vitro} and \textit{in vivo} frames and hence some debris is segmented as part of the stone, resulting in either underestimation or overestimation of the stone size.

\subsection{Generalizability study} 
\begin{figure}[t!]
    \centering
    \includegraphics[width=0.8\textwidth]{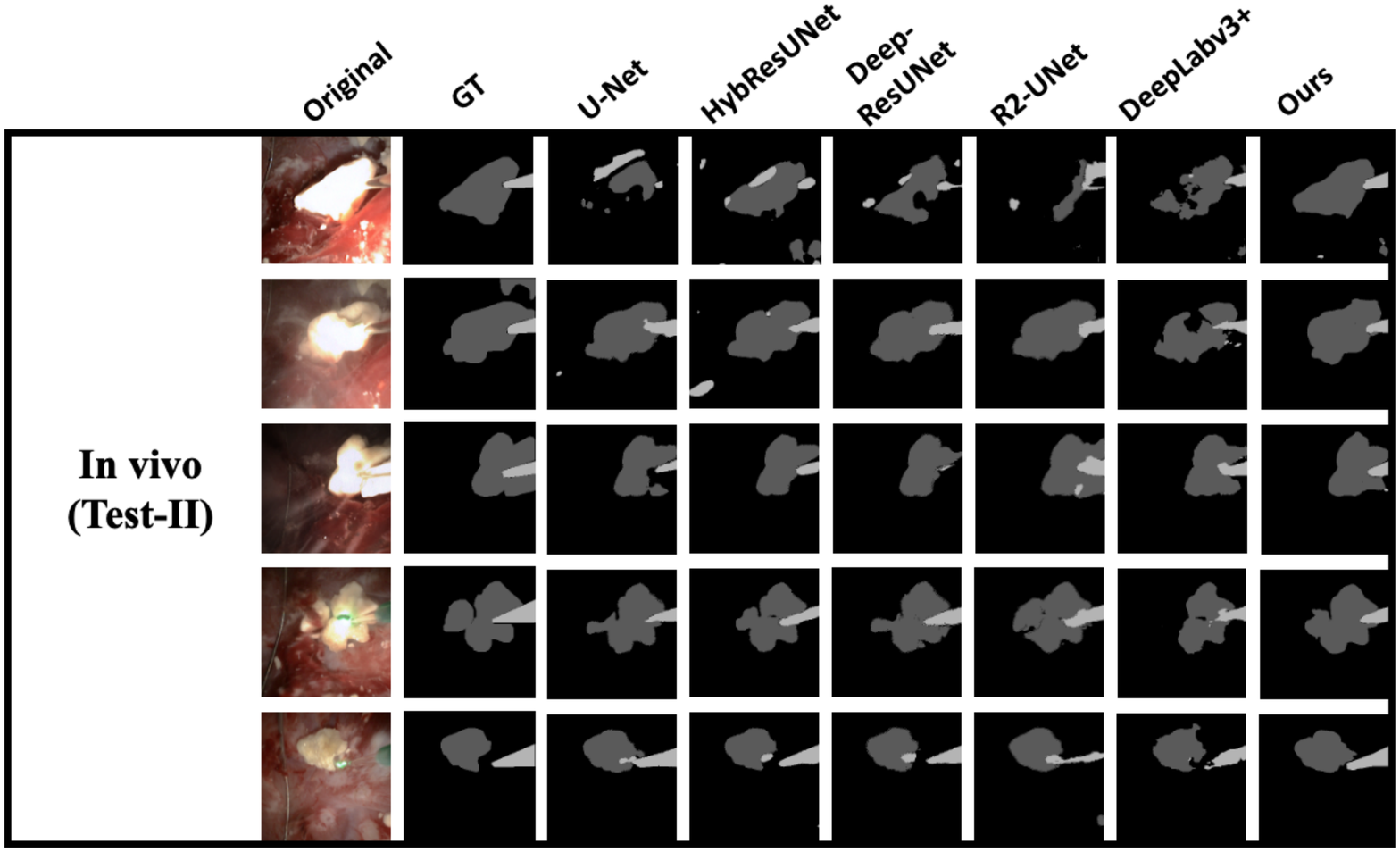}
    \caption{Qualitative analysis of our proposed  method  (HybResUNet+DVFNet (with warped image)) for \textit{in vivo} against existing SOTA methods methods on our unseen in the vivo test dataset (Test-II).}
    \label{qual_gen}
\end{figure}

We also performed an out-of-sample generalizability test of our proposed \textit{in vivo} framework for which an entirely new test dataset (Test-II) was used. We established our proposed method comparison with existing SOTA approaches. As shown in Table~\ref{quant_gen}, the proposed model (HybResUNet+DVFNet (with warped image)) provided a significant improvement with a mean (DSC and JI) of 25.4\%, 20\%, 23.2\%, 14.41\% and 11\% over UNet, HybResUNet, DeepResUNet,R2-UNet and DeepLabv3+, respectively. Qualitative results in Figure~\ref{qual_gen} on this dataset also demonstrate that our proposed model for the \textit{in vivo} is able to overcome the effect of debris and provide the most accurate segmentation of all frames as compared to the other existing approaches.

\begin{table}[]
\vspace{0.2in}
\small{
\begin{center}
\resizebox{\textwidth}{!}{%
\begin{tabular}{ |c|cc|c|c|c|c|c|c|c|c|c| }
\hline
\multirow{2}{*}{\bf Dataset} &
\multicolumn{2}{c|}{\multirow{2}{*}{Method}}&
\multicolumn{2}{c|}{\bf DSC}
&  \multicolumn{2}{c|}{\bf JI} &
 \multirow{1}{*}{\bf Mean} &
\multicolumn{2}{c|}{\bf PPV}
&  \multicolumn{2}{c|}{\bf Sensitivity} \\
\cline{4-7}

\cline{9-12}
& & & Stone & Laser  & Stone & Laser& \bf{(DSC, JI)} & Stone & Laser & Stone & Laser\\
\hline 
\multirow{5}{*}{\bf \textit{In vivo}}&&UNet~\cite{ronneberger2015u}  &	0.3516	&	0.5564	&	0.2382	&	0.4385	&	0.3962	&	0.4192	&	0.5824	&	0.3426	&	0.5746	\\

&& HybResUNet~\cite{peretz_amar_2019} &	0.3471	&	0.6003	&	0.2413	&	0.4673	&	0.4140	&	0.3977	&	0.6443	&	0.3451	&	0.5969	\\

 && DeepResUNet~\cite{zhang2018road} &	0.3558	&	0.5635	&	0.2471	&	0.4465	&	0.4032	&	0.4071	&	0.6255	&	0.3518	&	0.5452	\\

(Test-II)& &{R2-UNet}~\cite{alom2018recurrent} &	0.3508	&	0.6403	&	0.2406	&	0.505	&	0.4342	&	0.4221	&	0.686	&	0.3513	&	0.6417	\\

&& DeepLabv3+(Resnet-50)~\cite{Deeplabv3+}&	0.3662	&	0.6641	&	0.252	&	0.5071	&	0.4474	&	\bf{0.4288}	&	0.8078	&	0.348	&	0.5854	\\

&& \bf{HybResUNet+DVFNet (with warped image)}\textsuperscript{\textdagger} &	\bf{0.3854}	&	\bf{0.7333}	&	\bf{0.2728}	&	\bf{0.5956}	&	\bf{0.4968}	&	0.4262	&	\bf{0.8131}	&	\bf{0.3873}	&	\bf{0.7057}	\\

\hline

\end{tabular}
}
\end{center}
}
\caption{Quantitative comparison of proposed network architecture for \textit{in vivo} data against existing approaches on an unseen \textit{in vivo} test dataset (Test-II) \label{quant_gen}}
\end{table}

\subsection{Inference time analysis} 
The DVFNet part of the framework is only used for training the network and not used during inference time as indicated in Figure~\ref{network}. This allows the network to perform inference in real-time. In this section, we have shown a quantitative comparison of computation time analysis of our proposed base network (HybResUNet) against other base networks involved in this study and the results are presented in Table~\ref{time}. The inference time was calculated by running 10 test images on NVIDIA Quadro RTX 6000.  

\begin{table}[]
\vspace{0.2in}
\small{
\begin{center}
\begin{tabular}{ |c|c| } 
\hline
{\bf Network} & {\bf Computation time (secs)}\\
\hline
UNet~\cite{ronneberger2015u} & 0.8427 \\
HybResUNet~\cite{peretz_amar_2019} & 0.8623 \\
DeepResUNet~\cite{zhang2018road} & 0.8736 \\
R2-UNet~\cite{alom2018recurrent} & 1.2164 \\
DeepLabv3+(resnet50)~\cite{Deeplabv3+} & 0.7440 \\

\hline
\end{tabular}
\end{center}
}
\caption{Inference time of base network for 10 images of test set on NVIDIA Quadro RTX 6000 \label{time}}
\end{table}

%
\section*{Discussion}
Ureteroscopy with laser lithotripsy has evolved as the most commonly used procedure for stone fragmentation. Automated segmentation of kidney stones and laser fiber is an important initial step to performing any automated quantitative analysis of the stones, particularly stone-size estimation. Stone size is one of the most important parameters that is used by the surgeon to decide if the stone is small enough to be removed or requires further fragmentation. To date, not much work on automated segmentation of ureteroscopy and laser lithotripsy has been reported. This can be due to numerous inevitable challenges such as stone debris, dynamic background, intra-operative bleeding, varying illumination conditions, high variance in the appearance of stones, motion blur, specularities and other image artefacts~\cite{de2019complications,rosa2011algorithm}. To our knowledge this study is the first attempt to develop multi-class segmentation approach for ureteroscopy and laser lithotripsy. In this work, we conducted experiments for \textit{in vitro} and \textit{in vivo} datasets independently. This is due to the large variability that exists between the two datasets, as evident from Figure~\ref{size_dis}. The variability between the \textit{in vitro} and in the vivo datasets can also be seen from Figure~\ref{dataset}, which shows that the \textit{in vivo} background has more textural information, higher heterogeneity, debris from tissue and blood, and other image artefacts that are almost absent in the \textit{in vitro} dataset.

We first trained a HybResUNet~\cite{peretz_amar_2019} on our \textit{in vitro} and \textit{in vivo} datasets by using different augmentation techniques and recorded the DSC for each class to determine the optimized augmentation strategy that best captures our target data (refer to the \textbf{Supplementary material Table~\ref{aug}}). It can be observed from Figure~\ref{dataset} that the laser fiber is always present in a certain orientation and in the right part of the image. This explains why networks trained with spatial-level transforms such as flips, shift and rotate are not able to perform well on the test dataset. It can also be seen from \textbf{Supplementary material Table~\ref{aug}} that the random brightness contrast (RBC) and the histogram equalization in case of \textit{in vitro}, and RBC and contrast limited adaptive histogram equalization (CLAHE) in case of \textit{in vivo} dataset seem to either improve the segmentation accuracy or provide a competitive performance as compared to the case with no augmentations. Therefore, RBC+Equalize and RBC+CLAHE were used for data augmentation in all further experiments for \textit{in vitro} and \textit{in vivo} datasets, respectively. 

A compound loss function has been used in this study to enable better differentiation between easy or hard examples and improve the segmentation performance. In Table~\ref{lossfn}, we can see that the boundary loss when combined with focal loss leads to an improvement of 5.2\% and 9.5\% in case of the \textit{in vitro} and \textit{in vivo} datasets, respectively. The \textit{in vitro} dataset has less scope for improvement as it has no textural information in the background and relatively more pronounced boundaries for stone and laser class as compared to the \textit{in vivo} dataset. This explains the relatively higher improvement for the \textit{in vivo} as compared to the \textit{in vitro} dataset.
Incorporation of similarity loss and smoothness constraint further boosted the DSC for both \textit{in vitro} and \textit{in vivo} datasets.It can also be noted that the accuracy of laser class boosted by an overall higher margin as compared to the stone class. We hypothesize that such discrepancy in the improvement percentage between stone class and laser class is possibly due to the stone class having more variability in terms of shape and texture as compared to the laser class.

HybResUNet has a relatively simpler model with residual blocks in the encoder path only as opposed to the DeepResUNet~\cite{zhang2018road} and R2-UNet~\cite{alom2018recurrent}. We hypothesize that DeepResUNet and R2-UNet make the U-Net network more complicated and tend to overfit the training data, thereby leading to a poor performance (refer to Table~\ref{quant_network}) as compared to HybResUNet. Atrous Spatial Pyramid Pooling (ASPP) technique uses a series of different dilation rates in parallel to capture multi-scale contextual information. The use of dilated convolutions in an increasing order followed by decreasing order helps better aggregation of local features and improves detection of small objects~\cite{hamaguchi2018dilation}. This explains the improvement in the Dice similarity coefficient and Jaccard index of the laser class for ASPP-HybResUNet in both \textit{in vitro} and \textit{in vivo} datasets (refer to the \textbf{Supplementary material Table~\ref{quant_network-supplem}}). It can also be observed that when attention gate (AG) model is integrated into our encoder-decoder HybResUNet network, it improved the segmentation of stone class and also improved the sensitivity by suppressing irrelevant responses in the network. However, when we integrating attention gate mechanism, dilated convolutions and ASPP, the networks tend to become overly complicated and hence we did not observe any improvement in the segmentation performance~(refer to the \textbf{Supplementary material Table~\ref{quant_network-supplem}}). To further improve the segmentation performance of our network, we integrated DVFNet into our framework.
As can be observed in Table~\ref{quant_network}, DVFNet (with DVF) and DVFNet (with warped image) when integrated with HybResUNet improved the performance in the \textit{in vitro} and \textit{in vivo} datasets, respectively. HybResUNet+DVFNet (with DVF) provided a significant improvement by increasing the overall DSC and JI mean from 0.8434 to 0.8527 in case of the \textit{in vitro} dataset. This network when combined with the attention gate model, Att-HybResUnet+DVFNet (with DVF), slightly improved the overall mean further to 0.8580. On the other hand, HybResUNet+DVFNet (with warped image) in case of the \textit{in vivo} can be seen to provide the best quantitative and qualitative performance out of all other experiments.

In Figure~\ref{qualitativeresults}, our proposed framework is able to outperform other existing networks by overcoming the effect of debris and other artefacts in case of both \textit{in vitro} and \textit{in vivo} datasets. The first image in the \textit{in vivo} dataset in Figure~\ref{qualitativeresults} gets one of the worst performing results across all networks. It can be observed that the small stone on the top part of the image is less visible due to relatively less light falling on it and is therefore not getting captured by any models. It can also be observed that the bright-red tissue protrusion in this image is being misclassified as a stone by all the networks. 
This shows that the networks are sensitive to any tissue protrusions in the images. In addition to this, it should also be observed in this image that our model is able to provide the most accurate segmentation of the laser fiber as opposed to the other networks . We hypothesize that the other models fail to segment the laser accurately due to the fact that it is transparent in appearance and the laser light is not activated making it difficult for the network to spot it. Further, the segmentation prediction on the second and third \textit{in vivo} images show that our model is not only able to overcome the effect of stone debris but also blood and white tissue debris. Based on the results of fourth \textit{in vivo} image, we hypothesize that our proposed model is able to successfully pick up secondary small stones and perform well in dynamic illumination situations. 

Finally, we tried testing our proposed \textit{in vivo} framework on new unseen images (Test-II) to validate its generalizability with quantitative results and qualitative results presented in Table~\ref{quant_gen} and Figure~\ref{qual_gen}, respectively. As shown in Table~\ref{quant_gen}, our proposed framework outperformed the existing SOTA approaches on Test-II samples, thereby proving that our proposed model not only outperforms other SOTA networks but also performs quite well on new data. Although we have tried to mimic real scenarios in our \textit{in vitro} dataset by using real kidney stones for fragmentation, yet it has not been possible to capture every aspect of the clinical settings such as tissue background, proper flow of irrigation fluid, movement constraints on stone fragments, blood debris and other image artefacts. This possibly explains why our results and proposed framework is different for both \textit{in vitro} and \textit{in vivo} datasets. 

Stone and laser localization and segmentation can assist clinical stone fragmenting procedure. While, the video frame rate of most ureteroscopy is 30 FPS, our proposed framework achieves 12 FPS. Though we require a combination of HybResUNet and DVFNet for training, only HybResUNet is used during test inference minimizing the required time for segmentation of laser and stone.

\section{Conclusion}
To the best of our knowledge, we are the first to present a novel method with comprehensive experimental analysis on multi-class segmentation of ureteroscopy and laser lithotripsy imaging dataset. The proposed framework effectively makes use of residual connections and motion information between adjacent frames to produce robust and reliable segmentation of renal stones and laser fiber in real-time. The qualitative and quantitative results demonstrate that our algorithm can efficiently tackle the compromised vision quality within the kidney, resulting in increased segmentation accuracy as compared to the existing state-of-the-art methods. 
Our approach makes effective use of the temporal information within five adjacent frames only to improve the segmentation results. Future research direction includes using different forms of recurrent neural networks (RNNs) in order to improve the temporal information and use it to further improve the segmentation results. 
Future work also includes quantitative assessment of the segmented stone fragments and laser fiber in order to help the clinician gain a better understanding of the target and improve patient outcome. Through this study, we also highlight that the clinical workload of endoscopists can be tackled by the development of medical image analysis tools that can shorten procedure time whilst improving diagnosis and therapy.

\section*{Acknowledgements}
We would like to thank Boston Scientific (BSc) for funding this project (Grant No: DFR04690). We would also like to thank Aditi Ray, Longquan Chen and Niraj Rauniyar from BSc for facilitating the equipment, data and guidance. SG, LG and BT are funded by BSC, SA and JR are supported by the NIHR Oxford BRC, and JR. The computational aspects of this research were funded from the NIHR Oxford BRC with additional support from the Wellcome Trust Core Award Grant Number 203141/Z/16/Z.

\section*{Author contributions statement}
SG prepared the dataset, conducted all experiments, and wrote most part of the paper with input from SA and JR. SA and BT helped in the validation of the annotations. Clinical (\textit{in vivo}) and laboratory (\textit{in vitro}) datasets were acquired with the help and support of LG and BT. All authors reviewed the manuscript, provided substantial input and agreed to submit this manuscript. 

\section*{Additional information}
\textbf{Competing interests} The author(s) declare no competing interests.
\section*{Disclaimer}
Some of the data was acquired by or on behalf of Boston Scientific. Data on file using LithoVue and/or prototype devices. Concept device or technology not available for sale. 
\bibliography{ms.bib}

\begin{center}
  \textbf{\LARGE Multi-class motion-based semantic segmentation\\ for ureteroscopy and laser lithotripsy}\\
  
  \textbf{\LARGE Supplementary Material} \\
  \bigskip
  Soumya Gupta,$^{1,*}$ Sharib Ali,$^{1,2}$ , Louise Goldsmith,$^3$,
  Ben Turney,$^3$ and Jens Rittscher,$^{1,*}$\\[.1cm]
  {\itshape ${}^1$Institute of Biomedical Engineering and Big Data Institute,\\ Department of Engineering Science, University of Oxford, Oxford, UK \\
  ${}^2$Oxford NIHR Biomedical Research Centre, Oxford, UK\\
  ${}^3$Department of Urology, The Churchill, Oxford University Hospitals NHS Trust, Oxford, UK\\}
  ${}^*$soumya.gupta@eng.ox.ac.uk and jens.rittscher@eng.ox.ac.uk\\
\end{center}

\setcounter{section}{0}
\setcounter{equation}{0}
\setcounter{figure}{0}
\setcounter{table}{0}
\renewcommand{\theequation}{S\arabic{equation}}
\renewcommand{\thefigure}{S\arabic{figure}}

\section{Optimal data augmentation strategy}
The list of augmentation techniques with their corresponding settings used in this study are mentioned below in Table~\ref{aug-settings}. 
\renewcommand{\thetable}{S\arabic{table}}
\begin{table}[t!h!]
\vspace{0.2in}
\small{
\begin{center}
\begin{tabular}{l|c} 
\hline
{\bf Augmentation type} & {\bf Settings} \\
\hline
Horizontal Flip & Probability of application (p) = 0.5 \\
Vertical Flip & Probability of application (p) = 0.5\\
Shift scale rotate (SSR) & Shift limit=0.0625, Scale limit=0.1, Rotate limit=45, p=0.5\\
Sharpen &  Range [0.2,0.5], p=0.5\\
Gaussian Blur (GB) & Gaussian filter with the size of kernel in the range [3,7]\\
Random Brightness Contrast (RBC) & Range [-0.2, 0.2], p=0.5\\
Equalize & Probability of application (p) = 0.5\\
Contrast Limited Adaptive Histogram Equalization (CLAHE) & Contrast clip limit = 4.0, p=0.5 \\
\hline
\end{tabular}
\end{center}
}
\caption{Augmentation settings used in this study for both \textit{in vitro} and \textit{in vivo} datasets \label{aug-settings}}
\end{table}
    
\noindent Table~\ref{aug} below presents our ablation study associated with determining the optimal data augmentation strategy for our dataset. It can be seen that the \textit{Random Brightness Contrast(RBC)} and \textit{Equalize} transformation improve the segmentation accuracy in case of \textit{in vitro} dataset as compared to no augmentation scenario. On the other hand, \textit{RBC} and \textit{CLAHE} both provide a higher DSC compared to no augmentation in case of the \textit{in vivo} dataset.
\begin{table}[h!]
\vspace{0.2in}
\small{
\begin{center}
\begin{tabular}{l|c|c|c|c|c|c} 
\hline
\multirow{3}{*}{\bf Augmentation type} & \multicolumn{6}{c}{\bf DSC}\\
\cline{2-7}
& \multicolumn{3}{c|}{\bf \textit{In vitro}} & \multicolumn{3}{c}{\bf \textit{In vivo}}\\
\cline{2-7}
& Stone & Laser & Average & Stone & Laser & Average\\
\hline
None & 0.8732	&	0.8438	&	0.8585	&	0.8319	&	0.7936	&	0.81275 \\
\hline
Horizontal Flip  & 0.8241	&	0.8309	&	0.8275	&	0.8203	&	0.7947	&	0.8075 \\
Vertical Flip  & 0.8179	&	0.8054	&	0.81165	&	0.8102	&	0.7194	&	0.7648 \\
Shift scale rotate (SSR) & 0.8505	&	0.8217	&	0.8361	&	0.8261	&	0.7875	&	0.8068\\
Sharpen & 0.849	&	0.8177	&	0.83335	&	0.8065	&	0.8001	&	0.8033 \\
Gaussian Blur (GB) &0.8629	&	0.8467	&	0.8548	&	0.8126	&	0.8072	&	0.8099 \\
\bf{Random Brightness Contrast (RBC)} &{0.8743}	&	\bf{0.8774}	&	\bf{0.87585}	&	0.8265	&	0.8168	&	{0.8217}\\
\bf{Equalize} & \bf{0.8895}	&	0.8257	&	{0.8576}	&	0.775	&	0.8217	&	0.7983 \\
\bf{CLAHE} & 0.8571	&	0.8251	&	0.8411	&	\bf{0.839}	&	\bf{0.832}	&	\bf{0.8355} \\
\hline
\end{tabular}
\end{center}
}
\caption{Comparison of augmentation techniques (applied on HybResUNet) for both \textit{in vitro} and \textit{in vivo} datasets \label{aug}}
\end{table}
\vspace{2cm}
\section{Ablation study of network design}
Table~\ref{quant_network-supplem} presents an ablation study for integration of various combinations of dilations, ASPP and attention gate in different networks. 

\begin{enumerate}[label=(\Roman*)]
    \item \textbf{Base network with ASPP}: Here, we compare the effect of dilated convolutions and ASPP module connected to each base network presented in (I) in Table~\ref{quant_network-supplem}. A shown below in Table:~\ref{quant_network-supplem}), it can be seen that in case of ASPP-HybResUNet, ASPP improves the DSC and JI mean of laser in both \textit{in vitro} and \textit{in vivo} case by approximately 0.7\% and 1.4\%, respectively, but does not provide any overall improvement as compared to only HybResUNet. It can also be seen that ASPP increases the sensitivity for both stone and laser class but decreases the PPV of stone class in most networks for the \textit{in vitro} case. 
    
    \item \textbf{Base network with Attention(Att)}: In the second set of experiments in Table~\ref{quant_network-supplem}, we can see that using attention slightly improves the mean of DSC and JI for stone in case of \textit{in vitro} by 0.32\% and laser in case of the \textit{in vivo} by approximately 0.7\% for Att-HybResUNet. It can also be seen that attention increases the overall sensitivity of most networks.
    
    \item \textbf{Base network with Att-ASPP}: In our third set of experiments, we can see that the combination of both Attention and ASPP does not lead to any improvement in the segmentation of stone or laser in both \textit{in vitro} and \textit{in vivo} cases.
\end{enumerate}

\begin{table}[h!]
\vspace{0.2in}
\small{
\begin{center}
\resizebox{\textwidth}{!}{%
\begin{tabular}{|l|ll|c|c|c|c|c|c|c|c|c| }
\hline
\multirow{2}{*}{\bf Dataset} &
\multicolumn{2}{c|}{\multirow{2}{*}{\textbf{Method}}}&
\multicolumn{2}{c|}{\bf DSC}
&  \multicolumn{2}{c|}{\bf JI} &
 \multirow{1}{*}{\bf Mean} &
\multicolumn{2}{c|}{\bf PPV}
&  \multicolumn{2}{c|}{\bf Sensitivity} \\
\cline{4-7}

\cline{9-12}
& & & Stone & Laser  & Stone & Laser& \bf{(DC, JI)} & Stone & Laser & Stone & Laser\\
\hline 
\multirow{12}{*}{\bf \textit {In vitro}}&\multirow{4}{*}{\bf \text{I}}
&ASPP-UNet\textsuperscript{\textdagger}& 0.8504	&	0.8656	&	0.7580	&	0.7958	&	0.8175	&	0.8845	&	0.8612	&	0.8553	&	0.8780 \\
&&ASPP-HybResUNet\textsuperscript{\textdagger} & 0.8777	&	0.8794	&	0.8015	&	0.8075	&	0.8415	&	0.8917	&	0.8735	&	0.8997	&	0.9141 \\
&&ASPP-DeepResUNet\textsuperscript{\textdagger}&0.7934	&	0.8647	&	0.6825	&	0.7891	&	0.7824	&	0.7611	&	0.8537	&	0.9094	&	0.9006\\
&&ASPP-R2-UNet\textsuperscript{\textdagger} & 0.8506	&	{0.8832}	&	0.7597	&	{0.8106}	&	0.8260	&	0.8785	&	0.8717	&	0.8654	&	0.9192\\
\cline{2-12}
&\multirow{4}{*}{\bf \text{II}}&
Att-UNet~\cite{oktay2018attention} &	0.8730	&	0.8488	&	0.7925	&	0.7777	&	0.8230	&	0.9021	&	0.8426	&	0.8796	&	0.8657	\\
\multirow{4}{*}{(Test-I)}&& Att-HybResUNet\textsuperscript{\textdagger}&	 {0.8844}	&	0.8264	&	0.8192	&	0.7592	&	0.8223	&	0.9402	&	0.8402	&	0.8730	&	0.8168	\\
&&Att-DeepResUNet\textsuperscript{\textdagger}&	{0.8960}	&	0.8491	&	{0.8244}	&	0.7729	&	0.8356	&	0.9392	&	0.8404	&	0.8776	&	0.8746	\\
&&Att-R2-UNet~\cite{leejunhyun2019} &0.8788	&	0.8481	&	0.8007	&	0.7731	&	0.8252	&	0.9118	&	0.8291	&	0.8779	&	0.8814\\

\cline{2-12}
&\multirow{4}{*}{\bf \text{III}}
&Att-ASPP-UNet\textsuperscript{\textdagger} & 0.8301	&	0.8643	&	0.7357	&	0.7852	&	0.8038	&	0.8340	&	0.8615	&	0.8733	&	0.8807\\
&&Att-ASPP-HybResUNet\textsuperscript{\textdagger}&0.8641	&	0.8693	&	0.7790	&	0.8038	&	0.8291	&	0.8339	&	0.8794	&	\bf{0.9350}	&	0.8788\\
&&Att-ASPP-DeepResUNet\textsuperscript{\textdagger}& 0.8571	&	0.8578	&	0.7648	&	0.7775	&	0.8143	&	0.8788	&	0.8430	&	0.8635	&	0.9025 \\
&&{Att-ASPP-R2-UNet}\textsuperscript{\textdagger}&0.8338	&	0.8782	&	0.7370	&	0.8056	&	0.8137	&	0.8292	&	0.8577	&	0.8832	&	\bf{0.9340} \\
\cline{2-12}

\hline
\hline

\cline{2-12}
\multirow{12}{*}{\bf \textit{In vivo}}&\multirow{4}{*}{\bf \text{I}}&
ASPP-UNet\textsuperscript{\textdagger}& 0.7963	&	0.7865	&	0.6854	&	0.6928	&	0.7403	&	0.7991	&	0.7834	&	0.8416	&	0.801 \\
&&ASPP-HybResUNet\textsuperscript{\textdagger} & 0.8035	&	0.8334	&	0.6939	&	0.7453	&	0.7690	&	0.814	&	0.8249	&	0.833	&	0.8558 \\
&&ASPP-DeepResUNet\textsuperscript{\textdagger}&0.8234	&	0.7968	&	0.7152	&	0.699	&	0.7586	&	0.8282	&	0.816	&	0.861	&	0.7855\\
&&ASPP-R2-UNet\textsuperscript{\textdagger} & 0.7939	&	0.7409	&	0.6805	&	0.6453	&	0.7152	&	0.7986	&	0.7261	&	0.8334	&	0.7638\\
\cline{2-12}
&\multirow{4}{*}{\bf \text{II}}&
Att-UNet~\cite{oktay2018attention}&	0.8255	&	0.8094	&	0.7157	&	0.7292	&	0.7700	&	\bf{0.8526}	&	0.8325	&	0.8358	&	0.8316	\\
\multirow{4}{*}{(Test-I)}&& Att-HybResUNet\textsuperscript{\textdagger}&	0.819	&	0.8288	&	0.7177	&	0.7388	&	0.7761	&	0.8033	&	0.8763	&	0.8725	&	0.8231	\\
&&Att-DeepResUNet\textsuperscript{\textdagger}&	0.8119	&	0.7726	&	0.6994	&	0.6832	&	0.7418	&	0.8334	&	0.7736	&	0.8141	&	0.7825	\\
&&Att-R2-UNet~\cite{leejunhyun2019}&0.7985	&	0.6966	&	0.6794	&	0.5858	&	0.6901	&	0.7902	&	0.7343	&	0.8293	&	0.6893\\

\cline{2-12}
&\multirow{4}{*}{\bf \text{III}}
&Att-ASPP-UNet\textsuperscript{\textdagger} & 0.8079	&	{0.8364}	&	0.7019	&	{0.7478}	&	0.7735	&	0.7923	&	0.8222	&	0.8712	&	0.8606\\
&&Att-ASPP-HybResUNet\textsuperscript{\textdagger}&0.803	&	0.8229	&	0.6842	&	0.7282	&	0.7596	&	0.8391	&	0.8119	&	0.7982	&	0.8495\\
&&Att-ASPP-DeepResUNet\textsuperscript{\textdagger}& 0.8185	&	0.8087	&	0.712	&	0.7071	&	0.7616	&	0.8025	&	0.7999	&	0.87	&	0.8332 \\
&&{Att-ASPP-R2-UNet}\textsuperscript{\textdagger}&0.7814	&	0.7913	&	0.6617	&	0.6794	&	0.7285	&	0.7837	&	0.7943	&	0.8278	&	0.8032 \\
\cline{2-12}
\hline
\end{tabular}
}
\end{center}
}
\caption{Quantitative comparison of proposed network architectures against existing approaches on our ureteroscopy and laser lithotripsy test sets (Test-I). The published networks have been accordingly cited and our experiment networks have been indicated by a \textdagger superscript. \label{quant_network-supplem}}
\end{table}

\end{document}